\documentclass[12pt]{article}    % Specifies the document style.
\usepackage[ansinew]{inputenc} % permite utilizar los acentos y la ñ con el teclado directamente
\usepackage{hyperref}
\usepackage{graphics}
\usepackage{epsfig}
\usepackage{epstopdf}
\usepackage{graphicx} %for graphics inclusion
\usepackage{amssymb}  %provides various useful mathematical symbols
\usepackage{amsmath}
\usepackage{multirow} %provides the command needed for spanning rows
\usepackage{enumitem} % leftmargin - itemize
\usepackage{color}     % para colorear el texto
\usepackage{indentfirst}  % para sangran el primer párrafo de cada sección
\usepackage{booktabs}
\usepackage{caption} %configura la tabla y la figura
\usepackage{appendix}
\renewcommand\appendix{\section*{Appendix}}
\usepackage[gen]{eurosym}
\usepackage[scientific-notation=true]{siunitx}
\usepackage{epstopdf}

\usepackage[all]{xy}
\usepackage{bm}

\begin{document}
%%%%%%%%%%%%%%%%%%%%%%%%%%%%%%%%%%%%%
\newtheorem{definition}{Definition}
\newtheorem{theorem}{Theorem}
\newtheorem{example}{Example}
\newtheorem{corollary}{Corollary}
\newtheorem{lemma}{Lemma}
\newtheorem{proposition}{Proposition}
\newtheorem{remark}{Remark}
\newenvironment{proof}{{\bf Proof:\ \ }}{\qed}
\newcommand{\qed}{\rule{0.5em}{1.5ex}}
\newcommand{\bfg}[1]{\mbox{\boldmath $#1$\unboldmath}}

\begin{center}

\section*{The nonlinear distribution of employment across municipalities}

\vskip 0.2in {\sc \bf Faustino Prieto$^a$
\footnote{Corresponding author.\\
E-mail addresses: faustino.prieto@unican.es (F. Prieto),
sarabiaj@unican.es (J.M. Sarabia),
enrique.calderin@unimelb.edu.au (E. Calder\'{\i}n-Ojeda).}, 
Jos\'e Mar\'{\i}a Sarabia$^a$, Enrique Calder\'{\i}n-Ojeda$^b$
%\vskip 0.2in {\sc \bf Authors
\vskip 0.2in

{\small\it
$^a$Department of Economics, University of Cantabria, Avenida de los Castros s/n, 39005 Santander, Spain.\\
$^b$Centre for Actuarial Studies, Department of Economics, University of Melbourne, Melbourne, Australia.
}\\
}

\end{center}

\begin{abstract}\noindent
In this paper, the nonlinear distribution of employment across Spanish municipalities is analyzed.
In addition, we explore the properties of the family of generalized power law (GPL) distributions,
and test its adequacy for modelling employment data.
The hierarchical structure of the GPL family that includes
the hierarchy of Pareto (power law) distributions is deeply studied.
A new subfamily of heavy-tailed GPL distributions that is right tail equivalent
to a Pareto (power law) model is derived. Our findings show 
on the one hand that the distribution of employment across Spanish municipalities follows a power law
behavior in the upper tail and, on the other hand, the adequacy of GPL models for modelling employment 
data in the whole range of the distribution.
\end{abstract}
\vskip 0.2in

\noindent {\bf Key Words}: Spatial distribution of employment; Municipalities;
Generalized power law models; Complex Systems.

\section{Introduction}
The spatial distribution of employment has been analyzed in the economic literature from different angles. 
For instance, changes in the spatial concentration of employment at county level in service and non-service sectors
in US have been discussed in \cite{Desmet2005};
the effects of the labour market size on the wage inequality in \cite{Korpi2007};
the relation between employment density, industrial specialization and wage distribution is examined in \cite{Matano2011};
also, the relation between the development of the British and the German financial sector
and the spatial distribution of financial employment in those countries in \cite{Wojcik2015};
and the socio-economic impacts of the geographic labour force mobility in
Australia \cite{McKenzie2016,Rampellini2016};
among other references.

However, to the best of our knowledge, no papers have empirically discussed the spatial distribution of employment
(number of workers), at municipal level, in the whole range, from an statistical perspective - in contrast to, for example, the efforts
made to model the firm size distribution, in terms of number of workers \cite{Axtell2001}.

The aim of this study is twofold. Firstly, to determine which statistical
distribution is useful to model the spatial distribution of employment of Spanish municipalities,
observed on monthly basis, over the period from January 2003 to December 2017. For this purpose,
we have used freely available datasets available at {\tt www.seg-social.es} \cite{SS2018}. 
Secondly, a deeper exploration of the family of GPL probabilistic models and its applications is carried out. 
This family was firstly introduced by \cite{Prieto2017} as an extension of the power law models where the 
non-negative shape parameter is assumed to be expressible as a non-linear function of the data.

The first step of our analysis is based on testing the power law behavior of employment data in the upper tail.
Then, we use the family of generalized power law (GPL) distributions and
other well-known size distributions: Dagum, Lognormal, Lomax, Burr type XII (Singh-Maddala)
and Fisk (Log-logistic) distributions, to examine the whole range of the empirical distribution. 
Parameter estimation 
is performed by the method of maximum likelihood method. Next, model selection is conducted in terms 
of the Bayesian
information criterion, graphical validation (i.e. rank-size plots) and also by bootstrap resampling.

The rest of this paper is organized as follows: in Section \ref{se2}, we study the properties
of the family of Generalized Power Law (GPL) distributions, its connection with the Extreme Value Theory,
and its hierarchical structure that nests the hierarchy of Pareto (power law) distributions;
in Section \ref{se3}, we analyze the nonlinear distribution of employment across municipalities in Spain,
by providing empirical evidence of its power law behavior in the upper tail, and also of the efficacy 
of the family of GPL distributions to explain employment data
at municipal level in the whole range of the distribution; finally, conclusions are given in Section \ref{se4}.

\section{The family of generalized power law (GPL) distributions}\label{se2}
In this section, we firstly analyze the family of standard GPL distributions, firstly introduced in \cite{Prieto2017}. Next,
we find a new subfamily of regularly varying distributions at infinity, that belongs to the Maximum Domain
of attraction of the Fr\'echet distribution. The members of this heavy-tailed subfamily are right tail equivalent to a 
Pareto (power law) distribution. Some examples of members of the family of standard GPL distributions that belong and 
does not belong the new subfamily are given. Later, we examine the associated location-scale family of GPL distributions and provide
some examples of those models. Finally, we show that the family of GPL distributions admits a hierarchical
structure that nests the hierarchy of Pareto (power law) distributions as a particular case.

\subsection{The family of standard GPL distributions}
\begin{definition}
A continuous random variable $Z$ has a standard generalized power law (GPL) distribution if and only if
its cumulative distribution function (cdf) is given by
\begin{equation}\label{eq1}
F_Z(z)=\Pr(Z\leq z)=1-(1+z)^{-g(z)},\;z>0,
\end{equation}
and $F_Z(z)=0$ if $z\leq0$, where the real function $g:(0,\infty)\rightarrow\mathbb{R}^+$ is continuous,
positive, differentiable on $(0,\infty)$, and satisfies the following conditions
\begin{equation}\label{eq2}
\displaystyle\lim_{z \to 0^+}(1+z)^{g(z)}=1\;
\mbox{;}\;\displaystyle\lim_{z \to\infty}(1+z)^{g(z)}=\infty,
\end{equation}
and
\begin{equation}\label{eq3}
\displaystyle\frac{g'(z)}{g(z)}\geq\displaystyle\frac{-1}{(1+z)\log(1+z)},\;\forall\;z>0.
\end{equation}
where $g'(z)=d[g(z)]/dz$.
\end{definition}
It can be noted that the cdf (\ref{eq1}), satisfying conditions (\ref{eq2}) and (\ref{eq3}), is a genuine cdf, that it
can also be expressed as follows
\begin{equation*}\label{eq3bisbis}
F_Z(z)=\Pr(Z\leq z)= 1-\exp[-g(z)\log(1+z)],\;\forall\;z>0,
\end{equation*}
and that condition (\ref{eq3}) is equivalent to
\begin{equation*}\label{eq3bis}
\displaystyle\frac{d}{dz}\log[g(z)]\geq-\displaystyle\frac{d}{dz}\log[\log(1+z)],\;\forall\;z>0.
\end{equation*}
The survival function $S_Z(z)=1-F_Z(z)$ is given by
\begin{equation*}\label{eq4}
S_Z(z)=\Pr(Z>z)=(1+z)^{-g(z)},\;z>0,
\end{equation*}
and $S_Z(z)=1$ if $z\leq0$. The probability density function (pdf) is
\begin{equation*}\label{eq5}
f_Z(z)=\displaystyle\frac{d F_Z(z)}{dz}=\left[\displaystyle\frac{g(z)}{(1+z)}+g'(z)\log(1+z)\right](1+z)^{-g(z)}, \;z>0
\end{equation*}
and $f_Z(z)=0$ if $z\leq0$. In this paper, we show that the new subfamily of standard GPL distributions, defined
by the condition
\begin{equation}\label{eq6}
\displaystyle\lim_{z \to \infty}g(z)=\alpha>0,
\end{equation}
is comprised of regularly varying distributions at infinity, that belong to the Maximum Domain of Attraction of
the Fr\'echet distribution $MDA(\Phi_{\alpha})$ with $\Phi_{\alpha}=\exp\{-x^{-\alpha}\},x>0,\alpha>0$,
\cite{Ron1989, Ron2005, Seneta2006, Resnick2013}.
\begin{theorem}\label{th1}
Let us consider the family of standard GPL distributions defined in terms of the cdf (\ref{eq1}),
where the real function $g:(0,\infty)\rightarrow\mathbb{R}^+$ is continuous, positive, differentiable on $(0,\infty)$
and satisfies conditions (\ref{eq2}) and (\ref{eq3}).
Then, for any function $g$ with a strictly positive and finite limit at infinity,
$\displaystyle\lim_{z \to \infty}g(z)=\alpha>0$, the corresponding standard GPL distribution belongs
to the Maximum Domain of attraction of
the Fr\'echet distribution, that means that it is regularly varying at infinity with index $-\alpha<0$.
\end{theorem}
\begin{proof} See Appendix.
\end{proof}
\begin{remark}\label{rm1}
Let us consider the family of standard GPL distributions defined in terms of the cdf (\ref{eq1}),
where the real function $g:(0,\infty)\rightarrow\mathbb{R}^+$ is continuous, positive, differentiable on $(0,\infty)$
and satisfies conditions (\ref{eq2}), (\ref{eq3}) and (\ref{eq6}). Then, the corresponding standard GPL distribution
is right tail equivalent to a Pareto (power law) distribution and it is a heavy-tailed distribution.\\[-3ex]
\end{remark}
\begin{proof} See Appendix.
\end{proof}

Table \ref{tab1} shows some examples of standard GPL distributions satisfying condition (\ref{eq6}):
$\displaystyle\lim_{z \to \infty}g(z)=\alpha>0$. It is observable that Pareto distribution belongs to that subfamily
and corresponds to the choice $g(z)=\alpha$.  For the other examples (see \cite{Prieto2017}), we have
chosen only two-parameter models $\bm{\theta}^\mathsf{T}=(\alpha,\beta)$, that include Pareto distribution
as particular case when $\beta=0$.

\begin{table}[htp]\footnotesize
\renewcommand{\tablename}{\footnotesize{Table}}
\caption{\label{tab1}\footnotesize Some examples of standard GPL distributions with
$\displaystyle\lim_{z \to \infty}g(z)=\alpha>0$.
It can be noted that:
$S_Z(z)=\Pr (Z > z)=(1+z)^{-g(z)}, \forall z>0$; $S_Z(z)=1, \forall z\leq0$; and $f_Z(z)=0, \forall z\leq0$.}
\centering
\begin{tabular}{@{}l @{\hspace{0.5cm}} c  @{\hspace{0.5cm}} r @{}}
\toprule
$g(z)$, $\forall z>0$&
$\beta$&
$f_Z(z)=d F_z(z)/dz$, $\forall z>0$\\
\midrule
$\alpha$&
&
$\left[\displaystyle\frac{\alpha S_Z(z)}{1+z}\right]$\\[5mm]
$\alpha\left[1+\displaystyle\frac{\beta}{1+\log(1+z)}\right]$&
$\beta\geq-1$        &
$\left[1+\displaystyle\frac{\beta}{(1+\log(1+z))^2}\right]\left[\displaystyle\frac{\alpha S_Z(z)}{1+z}\right]$\\[5mm]
$\alpha\left[1+\displaystyle\frac{\beta z}{(1+z)\log(1+z)}\right]$&
$\beta\geq-1$&
$\left[1+\displaystyle\frac{\beta}{1+z}\right]\left[\displaystyle\frac{\alpha S_Z(z)}{1+z}\right]$\\[5mm]
$\alpha\left[\displaystyle\frac{z}{1+z}\right]^{\beta}$&
$\beta>-1$&
$\left[1+\displaystyle\frac{\beta\log(1+z)}{z}\right]\left[\displaystyle\frac{z}{1+z}\right]^{\beta}
\left[\displaystyle\frac{\alpha S_Z(z)}{1+z}\right]$\\[5mm]
$\alpha\left[\displaystyle\frac{\log (1+z)}{1+\log (1+z)}\right]^\beta$&
$\beta>-1$&
$\left[1+\displaystyle\frac{\beta}{1+\log(1+z)}\right]
\left[\displaystyle\frac{\log (1+z)}{1+\log (1+z)}\right]^\beta
\left[\displaystyle\frac{\alpha S_Z(z)}{1+z}\right]$\\[2ex]
\bottomrule
\end{tabular}
\end{table}

In contrast, Table \ref{tab2} illustrates some examples of standard GPL distributions with
$\displaystyle\lim_{z \to \infty}g(z)=\infty$. Thus, these models do not satisfy condition (\ref{eq6}). However, they do belong
to the family of GPL distributions, as they are defined in terms of the cdf (\ref{eq1}) and satisfy conditions
(\ref{eq2}) and (\ref{eq3}).
Again, only two-parameter models $\bm{\theta}^\mathsf{T}=(\alpha,\beta)$  that
include Pareto distribution as  particular case when $\beta=0$ have been considered.
The choice $g(z)=\alpha+\beta\log(1+z)$ corresponds to the
Benini distribution (\cite{Kleiber2003,Benini1905}) and the case $g(z)=\alpha\log^\beta(1+z)$ leads to the 
Pareto Positive Stable (PPS) distribution (\cite{Sarabia2009,Guillen2011}), in both cases the transformation $X=\sigma(1+Z)$ is needed.

\begin{table}[htp]\footnotesize
\renewcommand{\tablename}{\footnotesize{Table}}
\caption{\label{tab2}\footnotesize Some examples of standard GPL distributions with
$\displaystyle\lim_{z \to \infty}g(z)=\infty$, that include Pareto distribution as particular case when $\beta=0$.
($S_Z(z)=(1+z)^{-g(z)}, \forall z>0$; $\alpha>0$).}
\centering
\begin{tabular}{@{}l @{\hspace{0.55cm}} @{}c @{\hspace{0.55cm}} c  @{\hspace{0.55cm}} r @{}}
\toprule
$g(z)$, $\forall z>0$&
$\beta$&
Distribution&
$f_Z(z)=d F_z(z)/dz$, $\forall z>0$\\
\midrule
$\alpha[1+\beta(1+z)]$&
$\beta\geq0$&
&
$[1+\beta(1+z)(1+\log(1+z))]
\left[\displaystyle\frac{\alpha S_Z(z)}{1+z}\right]$\\[5mm]
$\alpha[1+\beta\log(1+z)]$&
$\beta\geq0$&
Benini&
$[1+2\beta\log(1+z)]
\left[\displaystyle\frac{\alpha S_Z(z)}{1+z}\right]$\\[5mm]
$\alpha\left[1+\displaystyle\frac{\beta z}{\log(1+z)}\right]$&
$\beta\geq0$&
&
$\left[1+\beta(1+z)\right]
\left[\displaystyle\frac{\alpha S_Z(z)}{1+z}\right]$\\[5mm]
$\alpha(1+z)^{\beta}$&
$\beta\geq0$&
&
$[1+\beta\log(1+z)](1+z)^{\beta}
\left[\displaystyle\frac{\alpha S_Z(z)}{1+z}\right]$\\[5mm]
$\alpha\log^\beta(1+z)$&
$\beta>-1$&
PPS&
$(1+\beta)\log^\beta(1+z)
\left[\displaystyle\frac{\alpha S_Z(z)}{1+z}\right]$\\[2ex]
\bottomrule
\end{tabular}
\end{table}

Obviously, both example lists of GPL models (tables \ref{tab1}, \ref{tab2}) can be easily 
expanded to more general models by increasing the number of parameters,
by not including Pareto model as  particular case, and/or by considering new forms of
the function $g$. For instance, by taking $\displaystyle\lim_{z \to \infty}g(z)=\infty$, the choice
$g(z)=z/\log(1+z)$ corresponds to the Exponential distribution,  $g(z)=z^2/[2\log(1+z)]$ coincides with the
Rayleigh distribution, the option $g(z)=z^\beta/\log(1+z)$ leads to the Weibull distribution,
and $g(z)=\beta(e^z-1)/\log(1+z)$ corresponds to the Gompertz distribution, after considering the
transformation $X=\sigma Z$. Table \ref{tab3} displays these well-known models, all members
of the family of generalized power law (GPL) distributions that do not include the Pareto
distribution as particular case.

\begin{table}[htp]\footnotesize
\renewcommand{\tablename}{\footnotesize{Table}}
\caption{\label{tab3}\footnotesize Other standard GPL distributions, with
$\displaystyle\lim_{z \to \infty}g(z)=\infty$, that do not include Pareto distribution
as particular case. ($S_Z(z)=(1+z)^{-g(z)}, \forall z>0$).}
\centering
\begin{tabular}{@{}l @{\hspace{1.6cm}} @{}c @{\hspace{1.6cm}} c  @{\hspace{1.6cm}} r @{}}
\toprule
$g(z)$, $\forall z>0$&
$\beta$&
Distribution&
$f_Z(z)=d F_z(z)/dz$, $\forall z>0$\\
\midrule
$\displaystyle\frac{z}{\log(1+z)}$&
&
Exponential&
$
S_Z(z)$\\[5mm]
$\displaystyle\frac{z^2}{2\log(1+z)}$&
&
Rayleigh&
$zS_Z(z)$\\[5mm]
$\displaystyle\frac{z^\beta}{\log(1+z)}$&
$\beta>0$&
Weibull&
$\beta z^{\beta-1}S_Z(z)$\\[5mm]
$\displaystyle\frac{\beta(e^z-1)}{\log(1+z)}$&
$\beta>0$&
Gompertz&
$\beta e^z S_Z(z)$\\[2ex]
\bottomrule
\end{tabular}
\end{table}

\subsection{A hierarchy of families of GPL models}
A location-scale family of GPL distribution can be defined related to the family of standard GPL distributions \cite{Prieto2017}:
\begin{definition}
Suppose that a random variable $Z$ has a standard generalized power law distribution defined by
expressions (\ref{eq1}),(\ref{eq2}),(\ref{eq3}).
Let $X=\mu+\sigma Z$, where $\mu \in \mathbb{R}$ is a location parameter  and $\sigma \in (0,\infty)$
is a scale parameter.
Then, $X$ belongs to the location-scale family of GPL distribution, with a cumulative distribution function (cdf)
given by
\begin{equation}\label{eq7}
F_X(x)=\Pr(X\leq x)=\left\{
\begin{array}{rl}
0,   & x\leq \mu \\
1-\left(1+\frac{x-\mu}{\sigma}\right)^{-g\left(\frac{x-\mu}{\sigma}\right)},  & x>\mu.
\end{array}
\right.
\end{equation}
\end{definition}
It is noted that the location parameter $\mu$ determines the lower bound of the support $(\mu,\infty)$ of
those models with cdf (\ref{eq7}), and that the corresponding survival function $S_X(x)=1-F_X(x)$ can be
expressed as
\begin{equation}\label{eq9}
S_X(x)=\left(1+\frac{x-\mu}{\sigma}\right)^{-g\left(\frac{x-\mu}{\sigma}\right)}, x>\mu
\end{equation}
and $S_X(x)=1$ if $x\leq \mu$. Table \ref{tab4} shows some examples of the location-scale family of
GPL distributions, with $\mu=0$ ($Z=X/\sigma$) and $\alpha,\sigma>0$, related to the examples
of standard GPL distributions exhibited in tables \ref{tab1}, \ref{tab2}, \ref{tab3}.

\begin{table}[htp]\footnotesize
\renewcommand{\tablename}{\footnotesize{Table}}
\caption{\label{tab4}\footnotesize
Some examples of location-scale GPL distributions pertaining to the examples of standard GPL
distributions illustrated in tables \ref{tab1}, \ref{tab2}, \ref{tab3}, with $\mu=0$ ($Z=X/\sigma$) and $\alpha,\sigma>0$.}
\centering
\begin{tabular}{@{}l @{\hspace{0.4cm}} @{}c @{\hspace{0.1cm}} c  @{\hspace{0.1cm}} r @{}}
\toprule
$g(z)$, $\forall z>0$&
$\beta$&
Distribution&
$S_X(x)=\Pr (X>x)$, $\forall x>0$\\
\midrule
$\alpha$&
&Pareto&
$[1+x/\sigma]^{-\alpha}$\\[3mm]
$\alpha\left[1+\displaystyle\frac{\beta}{1+\log(1+z)}\right]$&
$\beta\geq-1$&&
$[1+x/\sigma]^{-\alpha}
\exp{\left[-\alpha\beta\displaystyle\frac{\log(1+x/\sigma)}{1+\log(1+x/\sigma)}\right]}$\\[5mm]
$\alpha\left[1+\displaystyle\frac{\beta z}{(1+z)\log(1+z)}\right]$&
$\beta\geq-1$&&
$[1+x/\sigma]^{-\alpha}
\exp{\left[-\alpha\beta\displaystyle\frac{x/\sigma}{1+x/\sigma}\right]}$\\[5mm]
$\alpha\left[\displaystyle\frac{z}{1+z}\right]^{\beta}$&
$\beta>-1$&&
$\exp{\left[-\alpha\log(1+x/\sigma)\left(\displaystyle\frac{x/\sigma}{1+x/\sigma}\right)^\beta\right]}$\\[5mm]
$\alpha\left[\displaystyle\frac{\log (1+z)}{1+\log (1+z)}\right]^\beta$&
$\beta>-1$&&
$\exp{\left[-\alpha\displaystyle\frac{\log^{\beta+1}(1+x/\sigma)}{[1+\log(1+x/\sigma)]^\beta}\right]}$\\[5mm]
\midrule
$\alpha[1+\beta(1+z)]$&
$\beta\geq0$&
&
$[1+x/\sigma]^{-\alpha}
\exp{\left[-\alpha\beta(1+x/\sigma)\log(1+x/\sigma)\right]}$\\[5mm]
$\alpha[1+\beta\log(1+z)]$&
$\beta\geq0$&
Benini&
$[1+x/\sigma]^{-\alpha}
\exp{\left[-\alpha\beta\log^2(1+x/\sigma)\right]}$\\[5mm]
$\alpha\left[1+\displaystyle\frac{\beta z}{\log(1+z)}\right]$&
$\beta\geq0$&
&
$[1+x/\sigma]^{-\alpha}
\exp{\left[-\alpha\beta(x/\sigma)\right]}$\\[5mm]
$\alpha(1+z)^{\beta}$&
$\beta\geq0$&
&
$\exp{\left[-\alpha\log(1+x/\sigma)\left(1+x/\sigma\right)^\beta\right]}$\\[5mm]
$\alpha\log^\beta(1+z)$&
$\beta>-1$&
PPS&
$\exp{\left[-\alpha\log^{\beta+1}(1+x/\sigma)\right]}$\\[5mm]
\midrule
$\displaystyle\frac{z}{\log(1+z)}$&
&
Exponential&
$\exp{\left[-x/\sigma\right]}$\\[5mm]
$\displaystyle\frac{z^2}{2\log(1+z)}$&
&
Rayleigh&
$\exp{\left[-x^2/(2\sigma^2)\right]}$\\[5mm]
$\displaystyle\frac{z^\beta}{\log(1+z)}$&
$\beta>0$&
Weibull&
$\exp{\left[-(x/\sigma)^\beta\right]}$\\[5mm]
$\displaystyle\frac{\beta(e^z-1)}{\log(1+z)}$&
$\beta>0$&
Gompertz&
$\exp{\left[-\beta(e^{x/\sigma}-1)\right]}$\\[5mm]
\bottomrule
\end{tabular}
\end{table}

Next, we show that the family of GPL models admits a hierarchical structure that includes the
hierarchy of Pareto (power law) distributions (\cite{Arnold2008,Arnold2014,Arnold2015}) as particular case.

\subsubsection{$GPL(I)$ family of distributions}

As baseline model of the GPL hierarchy, we use the location-scale GPL model,
defined by cdf (\ref{eq7}),
with $\mu=\sigma>0$. We name this family as $GPL(I)$ family of distributions.  Their survival functions
have the following form
\begin{equation}\label{eq12}
S_X(x)=1-F_X(x)=\left(1+\frac{x-\sigma}{\sigma}\right)^{-g\left(\frac{x-\sigma}{\sigma}\right)}=
\left(\frac{x}{\sigma}\right)^{-g\left(\frac{x}{\sigma}-1\right)}, \;x>\sigma.
\end{equation}
A random variable $X$ with survival function given by Eq.(\ref{eq12}) will be denoted
by $X\sim GPL(I)(\sigma,\bm{\theta}^\mathsf{T})$,
where $\bm{\theta}^\mathsf{T}$ is the row vector of additional parameters of the function $g$.

By taking the constant function
$g\left(\frac{x-\sigma}{\sigma}\right)=\alpha>0$, we obtain the classical Pareto distribution,
denoted as $P(I)(\sigma,\alpha)$, with survival function given by
\begin{equation}\label{eq13}
S_X(x)=\left(\frac{x}{\sigma}\right)^{-\alpha}, \;x>\sigma.
\end{equation}

\subsubsection{$GPL(II)$ family of distributions}

A more general family of GPL distributions, denoted as $GPL(II)$, can be defined in terms
of survival function, coinciding with Eq.(\ref{eq9}),  as follows:
\begin{equation}\label{eq14}
S_X(x)=\left(1+\displaystyle\frac{x-\mu}{\sigma}\right)^{-g\left(\frac{x-\mu}{\sigma}\right)}, \;x>\mu,
\end{equation}
where $\mu \in \mathbb{R}$ is a location parameter and $\sigma>0$ is a scale parameter.
A random variable $X$ with survival function given by Eq.(\ref{eq14}) will be denoted
by $X\sim GPL(II)(\mu,\sigma,\bm{\theta}^\mathsf{T})$,
where $\bm{\theta}^\mathsf{T}$ is the row vector of additional parameters of the function $g$.

As a particular case, by again considering the constant function $g\left(\frac{x-\mu}{\sigma}\right)=\alpha>0$, we obtain
the well-known Pareto type II distribution, denoted as $P(II)(\mu,\sigma,\alpha)$, with survival function given by
\begin{equation*}\label{eq14bis}
S_X(x)=\left(1+\displaystyle\frac{x-\mu}{\sigma}\right)^{-\alpha}, \;x>\mu,
\end{equation*}
that includes Lomax distribution \cite{Lomax1954} when $\mu=0$,  with survival function
\begin{equation}\label{eq15bis}
S_X(x)=\left(1+\displaystyle\frac{x}{\sigma}\right)^{-\alpha}, \;x>0.
\end{equation}

\subsubsection{$GPL(III)$ family of distributions}

It is also possible to include an additional shape parameter $\gamma>0$ and generalize the family
of GPL distributions to a wider class, denoted as $GPL(III)$. This family can be defined in terms of the survival function
as follows
\begin{equation}\label{eq16}
S_X(x)=\left[1+\left(\displaystyle\frac{x-\mu}{\sigma}\right)^{1/\gamma}\right]^{-g\left[\left(\frac{x-\mu}{\sigma}\right)^{1/\gamma}\right]}, \;x>\mu,\\[-1ex]
\end{equation}
where $\mu \in \mathbb{R}$ is a location parameter, $\sigma>0$ is a scale parameter.

A random variable $X$ with survival function given by Eq.(\ref{eq16}) will be denoted by
$X\sim GPL(III)(\mu,\sigma,\gamma,\bm{\theta}^\mathsf{T})$,
where $\bm{\theta}^\mathsf{T}$ is the row vector of additional parameters of the function $g$.

We can obtain as particular cases the following models
\begin{itemize}[leftmargin=*]
\item By taking the constant function
$g\left[\left(\frac{x-\mu}{\sigma}\right)^{1/\gamma}\right]=1$, we obtain the Pareto type III distribution,
denoted as $P(III)(\mu,\sigma,\gamma)$, with survival function given by
\begin{equation*}\label{eq17}
S_X(x)=\left[1+\left(\displaystyle\frac{x-\mu}{\sigma}\right)^{1/\gamma}\right]^{-1}, \;x>\mu.
\end{equation*}
Note that the latter model includes the Fisk distribution, also known as Log-logistic distribution (\cite{Kleiber2003,Fisk1961}),
when $\mu=0$, with survival function,
\begin{equation*}\label{eq18bisbis}
S_X(x)=\left[1+\left(\displaystyle\frac{x}{\sigma}\right)^{1/\gamma}\right]^{-1}, \;x>0;
\end{equation*}

\item by taking the constant function
$g\left[\left(\frac{x-\mu}{\sigma}\right)^\gamma\right]=\alpha>0$, the Pareto type IV distribution,
denoted as $P(IV)(\mu,\sigma,\gamma,\alpha)$, is obtained with survival function given by
\begin{equation*}\label{eq18}
S_X(x)=\left[1+\left(\displaystyle\frac{x-\mu}{\sigma}\right)^{1/\gamma}\right]^{-\alpha}, \;x>\mu.
\end{equation*}
Note that this model includes the Burr type XII distribution, also known as Singh-Maddala
distribution (\cite{Burr1942,Singh1976,Kotz2004}), when $\mu=0$, with survival function
\begin{equation*}\label{eq18bis}
S_X(x)=\left[1+\left(\displaystyle\frac{x}{\sigma}\right)^{1/\gamma}\right]^{-\alpha}, \;x>0;
\end{equation*}
\end{itemize}

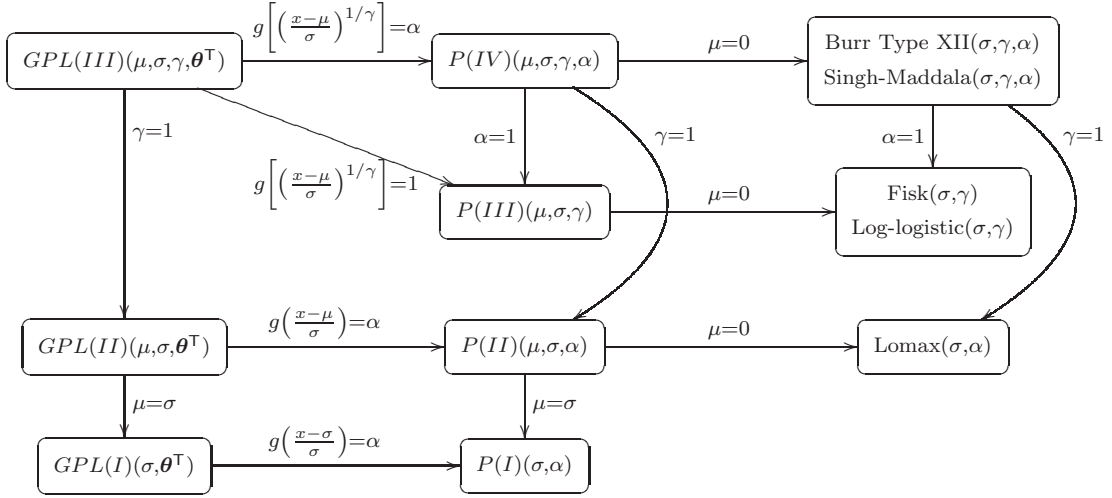
\begin{figure}[htp]
\renewcommand{\figurename}{\footnotesize{Figure}}
\begin{center}
$\def\objectstyle{\scriptstyle}
\def\labelstyle{\scriptstyle}
\vcenter
{\xymatrixcolsep{6pc}\xymatrix{
*++[F-:<3pt>]{GPL(III)(\mu,\sigma,\gamma,\bm{\theta}^\mathsf{T})} \ar[dd]^(0.25){\gamma=1}  \ar[r]^-{g\left[\left(\frac{x-\mu}{\sigma}\right)^{1/\gamma}\right]=\alpha}    \ar[rd]_(0.6){\;\;\;g\left[\left(\frac{x-\mu}{\sigma}\right)^{1/\gamma}\right]=1}
  &  *++[F-:<3pt>]{P(IV)(\mu,\sigma,\gamma,\alpha)} \ar[d]_(0.5){\alpha=1}
  \ar@/^4.5pc/[dd]^(0.31){\gamma=1}   \ar[r]^(0.5){\mu=0} &
   *++[F-:<3pt>]{\txt{\scriptsize Burr Type XII($\sigma$,$\gamma$,$\alpha$)\\ \scriptsize Singh-Maddala($\sigma$,$\gamma$,$\alpha$)  }} \ar[d]_(0.5){\alpha=1} \ar@/^4.5pc/[dd]^(0.31){\gamma=1}\\
   &  *++[F-:<3pt>]{P(III)(\mu,\sigma,\gamma)}   \ar[r]^(0.5){\mu=0}   &
   *++[F-:<3pt>]{\txt{\scriptsize Fisk($\sigma$,$\gamma$)\\ \scriptsize Log-logistic($\sigma$,$\gamma$)  }}\\
*++[F-:<3pt>]{GPL(II)(\mu,\sigma,\bm{\theta}^\mathsf{T})}    \ar[d]^{\mu=\sigma}
\ar[r]^(0.5){g\left(\frac{x-\mu}{\sigma}\right)=\alpha}
                   & *++[F-:<3pt>]{P(II)(\mu,\sigma,\alpha)}     \ar[d]^{\mu=\sigma}   \ar[r]^(0.5){\mu=0} & *++[F-:<3pt>]{\txt{\scriptsize Lomax($\sigma$,$\alpha$)}}\\
*++[F-:<3pt>]{GPL(I)(\sigma,\bm{\theta}^\mathsf{T})}  \ar[r]^(0.5){g\left(\frac{x-\sigma}{\sigma}\right)=\alpha}
                    &  *++[F-:<3pt>]{P(I)(\sigma,\alpha)}        &
}}$
\caption{\label{fig01}\footnotesize{Relationship between the family of GPL distributions, Pareto type distributions
and other common size distributions.}}
\end{center}
\end{figure}

Figure \ref{fig01} shows the relationship between the family of GPL distributions and the family of
Pareto-type distributions, their hierarchies, and between other common size distributions used in the literature. A solid line means an exact relationship (special case).
Table \ref{tab5} illustrates the corresponding survival function of those probabilistic families. In particular,
it is noted that the families $GPL(I)$ and $GPL(II)$ can be identified as special cases of the
$GPL(III)$ family as follows
\begin{eqnarray*}\label{eq19}
GPL(I)(\sigma,\bm{\theta}^\mathsf{T})&\equiv&GPL(III)(\sigma,\sigma,1,\bm{\theta}^\mathsf{T})\\
GPL(II)(\mu,\sigma,\bm{\theta}^\mathsf{T})&\equiv&GPL(III)(\mu,\sigma,1,\bm{\theta}^\mathsf{T})
\end{eqnarray*}

\begin{table}[htp]\scriptsize
\renewcommand{\tablename}{\footnotesize{Table}}
\caption{\label{tab5}\footnotesize Survival Function $S_X(x)=\Pr (X>x)$ of the GPL family of distributions,
Pareto type distributions and other common size distributions. ($S_X(x)=1, \forall x\leq\mu$) }
\centering
\begin{tabular}{@{}c @{\hspace{0.2cm}} c  @{\hspace{0.0cm}} c @{}}
\toprule
&
$P(IV)(\mu,\sigma,\gamma,\alpha)$&
Burr(XII)($\sigma$,$\gamma$,$\alpha$)$\equiv P(IV)(0,\sigma,\gamma,\alpha)$\\[2mm]

$GPL(III)(\mu,\sigma,\gamma,\bm{\theta}^\mathsf{T})$&
$S_X(x)=\left[1+\left(\displaystyle\frac{x-\mu}{\sigma}\right)^{1/\gamma}\right]^{-\alpha}$&
$S_X(x)=\left[1+\left(\displaystyle\frac{x}{\sigma}\right)^{1/\gamma}\right]^{-\alpha}$\\[3mm]

\multirow{2}{*}{$S_X(x)=\left[1+\left(\displaystyle\frac{x-\mu}{\sigma}\right)^{1/\gamma}\right]^{-g\left[\left(\frac{x-\mu}{\sigma}\right)^{1/\gamma}\right]}$}&
$x>\mu$&
$x>0$\\[2mm]

\cmidrule{2-3}

&
$P(III)(\mu,\sigma,\gamma)\equiv P(IV)(\mu,\sigma,\gamma,1)$&
Fisk($\sigma$,$\gamma$)$\equiv P(III)(0,\sigma,\gamma)$\\[2mm]

$x>\mu$&
$S_X(x)=\left[1+\left(\displaystyle\frac{x-\mu}{\sigma}\right)^{1/\gamma}\right]^{-1}$&
$S_X(x)=\left[1+\left(\displaystyle\frac{x}{\sigma}\right)^{1/\gamma}\right]^{-1}$\\[3mm]

&
$x>\mu$&
$x>0$\\[2mm]

\midrule

$GPL(II)(\mu,\sigma,\bm{\theta}^\mathsf{T})\equiv GPL(III)(\mu,\sigma,1,\bm{\theta}^\mathsf{T})$&
$P(II)(\mu,\sigma,\alpha)\equiv P(IV)(\mu,\sigma,1,\alpha)$&
Lomax($\sigma$,$\alpha$)$\equiv P(II)(0,\sigma,\alpha)$\\[2mm]

$S_X(x)=\left(1+\displaystyle\frac{x-\mu}{\sigma}\right)^{-g\left(\frac{x-\mu}{\sigma}\right)}$&
$S_X(x)=\left(1+\displaystyle\frac{x-\mu}{\sigma}\right)^{-\alpha}$&
$S_X(x)=\left(1+\displaystyle\frac{x}{\sigma}\right)^{-\alpha}$\\[3mm]

$x>\mu$&
$x>\mu$&
$x>0$\\[2mm]

\midrule

$GPL(I)(\sigma,\bm{\theta}^\mathsf{T})\equiv GPL(III)(\sigma,\sigma,1,\bm{\theta}^\mathsf{T})$&
$P(I)(\sigma,\alpha)\equiv P(IV)(\sigma,\sigma,1,\alpha)$&
\\[2mm]

$S_X(x)=\left(\displaystyle\frac{x}{\sigma}\right)^{-g\left(\frac{x-\sigma}{\sigma}\right)}$&
$S_X(x)=\left(\displaystyle\frac{x}{\sigma}\right)^{-\alpha}$&
\\[3mm]

$x>\sigma$&
$x>\sigma$&
\\[2mm]

\bottomrule
\end{tabular}
\end{table}

Finally, we can extend Theorem \ref{th1}, and Remark \ref{rm1}, to the hierarchy of GPL family of distributions, as follows:

\begin{remark}\label{rm2}
Let us consider the $GPL(I)$, $GPL(II)$, $GPL(III)$ families of distributions defined in terms
of the survival functions (\ref{eq12}),  (\ref{eq14}), (\ref{eq16}) respectively,
where the real function $g:(0,\infty)\rightarrow\mathbb{R}^+$ is continuous, positive, differentiable
on $(0,\infty)$ and satisfies conditions (\ref{eq2}) and (\ref{eq3}).
Then, for any function $g$ with a strictly positive and finite limit at infinity,
$\displaystyle\lim_{z \to \infty}g(z)=\alpha>0,$
the corresponding GPL distribution belongs to the Maximum Domain of attraction of the Fr\'echet distribution,
it is regularly varying at infinity with index $-\alpha<0$, it is right tail equivalent to a Pareto (power law)
distribution, and it is a heavy-tailed distribution.
\end{remark}
\begin{proof} See Appendix.
\end{proof}

\section{Nonlinear distribution of employment across Spanish municipalities}\label{se3}

In this Section, we analyze the nonlinear distribution of employment across municipalities in Spain,
in terms of the number of workers affiliated to the Spanish Social Security in each municipality. We show
that it can be modelled by means of a Generalized Power Law (GPL) distribution in the whole range. It is also illustrated that it follows
a power law behavior in the upper tail of the distribution.

\subsection{The Data}

Now we use employment data of Spanish municipalities related to the number of workers
affiliated to the Spanish Social Security system, in one of its general or special schemes (employed or
self-employed), in each municipality. Those data are available on monthly basis, from
the last day of each month, since January 2003, on the Spanish Social Security website,
{\tt www.seg-social.es} \cite{SS2018}. Our analysis comprises  180 different samples from January 2003 to December 2017. It should be noted that the data are left-censored: municipalities
with less than or equal to 4 workers are reported as  "$<5$".

\begin{figure}[p]
\renewcommand{\figurename}{\footnotesize{Figure}}
\begin{center}
\includegraphics[scale=0.67]{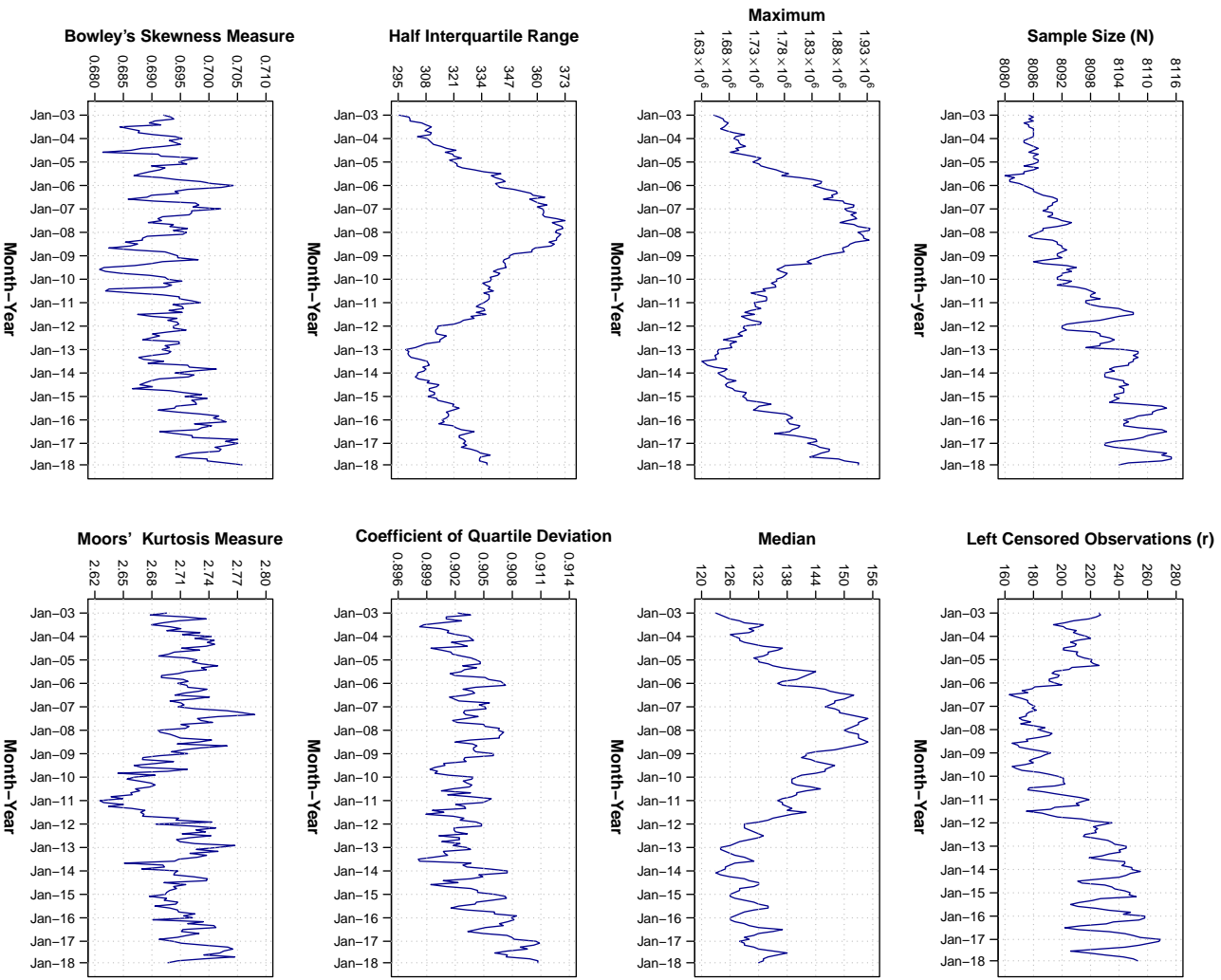}
\caption{\footnotesize{Plots of the sample size ($N$), number of left censored observations ($r$),
maximum value, median, half interquartile range, coefficient of quartile deviation, Bowley's skewness measure and Moors' kurtosis measure, for each of the 180 monthly samples
considered, from January 2003 to December 2017.}}\label{fig02}
\end{center}
\end{figure}

Figure \ref{fig02} shows, for each of the 180 different months considered:\\[-4ex]
\begin{itemize}[leftmargin=4mm]
\item the total number $N$ of municipalities (sample size for each month) - as a reference, there were
$N=8104$ Spanish municipalities in December 2017 (last month considered);\\[-4ex]
\item the total number $r$ of left-censored municipalities for each month (municipalities with 4 or less
workers affiliated, and reported as "$<5$") - for example, there were $r=253$ left-censored municipalities
in December 2017 (thus, there were 7851 not-left-censored municipalities, with 5 o more workers affiliated); \\[-4ex]
\item the maximum number of workers affiliated in a municipality for each month (corresponds to Madrid,
Spanish capital) - for example, there were 1,913,935 workers affiliated in Madrid in December 2017;\\[-4ex]
\item and some measures for location, dispersion, skewness and kurtosis, based on quantiles, for the
variable of interest: \\[-4ex]
\begin{itemize}[leftmargin=4mm]
\item Median $M=Q(0.50)$,
\item Half interquartile range $R=[Q(0.75)-Q(0.25)]/2$,
\item Coefficient of quartile deviation $=[Q(0.75)-Q(0.25)]/[Q(0.75)+Q(0.25)]$,
\item Bowley's \cite{Bowley1920} skewness measure\\ $S=[Q(0.75)-2Q(0.5)+Q(0.25)]/[Q(0.75)-Q(0.25)]$,
\item Moors' \cite{Moors1988} kurtosis measure\\  $T=[(Q(7/8)-Q(5/8))-(Q(3/8)-Q(1/8))]/[Q(6/8)-Q(2/8)]$.
\end{itemize}
\end{itemize}

\subsection{Analyzing power-law data}

A power law model (also known as classical Pareto distribution) can be defined, in terms of its survival function,
by Eq.(\ref{eq13}). That model is a member of the GPL family of distributions and it is also a particular case of other members of the same family (see Table \ref{tab4}).  Now, we show that the empirical distribution of employment across Spanish municipalities adheres to a power law in its upper tail.

For that reason, we followed the methodology proposed by Clauset et al. in (\cite{Clauset2009,Efron,Babu,Prieto2017}) which is based on the following steps: first, application of  the maximum likelihood estimation method, for fitting the Pareto distribution to the data; second, the utilization of the Kolmogorov-Smirnov test by using bootstrap resampling, for testing  the goodness-of-fit of that model; and finally the application of an iterative algorithm for estimating the lower bound of the power law behaviour,  which is the
minimum sample value where the null hypothesis {\it $H_0$: the data follow a power law model} cannot be rejected at 0.1 significance level.

\begin{figure}[p]
\renewcommand{\figurename}{\footnotesize{Figure}}
\begin{center}
\includegraphics[scale=0.67]{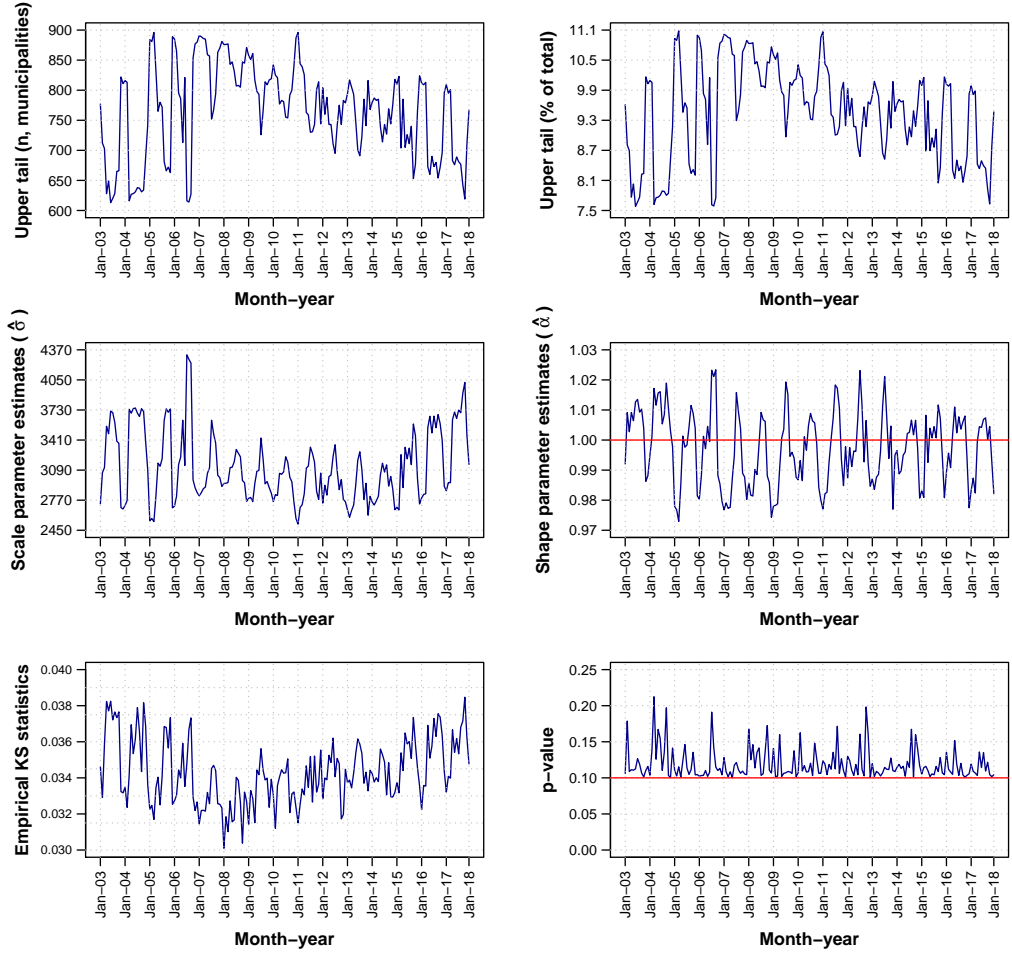}
\caption{\footnotesize{Plots of the size of the upper tail ($n$),
percentage of the size of the upper tail with respect to total,
scale parameter estimates $\hat{\sigma}$ (lower bound of the power law behavior),
shape parameter estimates $\hat{\alpha}$ (very close to 1, Zipf's law),
empirical Kolmogorov-Smirnov statistics ($KS$),
and the corresponding bootstrap $p$-values
(as the $p$-values exceed the proposed significance level of 0.1,
there is not statistical evidence to reject {\it$H_0$: the data follow a power law model}, in the upper tail
of the distribution),
in each of the 180 monthly samples considered,
from January 2003 to December 2017.
}}\label{fig03}
\end{center}
\end{figure}

Figure \ref{fig03} displays the obtained outcomes, for each of the 180 months considered: \\[-4ex]
\begin{itemize}[leftmargin=4mm]
\item the size of the upper tail $n$: number of largest municipalities whose employment follows a
power law behavior (where the null hypothesis, {\it$H_0$: the data follow a power law model},
cannot be rejected at 0.1 significance level).
For example, the $n=768$ largest Spanish municipalities followed a power law behavior in December 2017;\\[-4ex]
\item the percentage of the size of the upper tail with respect to the total number of municipalities: $100\;n/N$.
For instance, in December 2017, it represented a 9.48 \% of the total ($100\;n/N=100\times768/8104$), \\[-4ex]
\item the scale parameter estimates $\hat{\sigma}$, obtained by maximum likelihood, coincides with the
lower bound $x_{min}$ of the power law behavior, ($\hat{\sigma}=x_{min}$).
For example, in December 2017, the municipality in the lower bound of the power law behavior
had  3,144 workers;  \\[-4ex]
\item the shape parameter estimates again obtained by maximum likelihood,
i.e. Hill estimator \cite{Hill1975},
$\hat{\alpha}=n\left[\sum_{i=1}^n\log(x_i/x_{min})\right]^{-1}$.
It is observable that $\hat{\alpha}$ are very close to 1
in all the 180 months considered (in this case, power law is also known as Zipf's law \cite{Ioannides,Fujiwara,Gabaix2016});\\[-4ex]
\item the empirical Kolmogorov-Smirnov statistics
$KS=\sup\;\lvert F_n(x_{i})-F(x_{i};\hat{\alpha},\hat{\sigma}) \rvert,$ $i=1,\dots,n$,
where $F_n(x_{i})\approx(n+1)^{-1}\sum_{j=1}^n I_{[x_{j}\leq x_{i}]}$ is the empirical cdf in a sample value
with the indicated plotting position formula \cite{Castillo2005}, and
$F(x_{i};\hat{\alpha},\hat{\sigma})$ is the theoretical cumulative distribution function (cdf) of the Pareto
model fitted by maximum likelihood in a sample value; \\[-4ex]
\item the corresponding bootstrap $p$-values, obtained from 10,000 synthetic datasets (simulated by
using $\hat{\alpha},\hat{\sigma}$), and calculated as a fraction of the times that the empirical $KS$ value is
better (less) than the corresponding 10,000 simulated $KS$ statistics.
In all the 180 months considered, as the p-value exceeds the proposed significance level,
there is not statistical evidence to reject {\it$H_0$: the data follow a power law model}, in the upper
tail of the distribution.
\end{itemize}

\begin{figure}[p]
\renewcommand{\figurename}{\footnotesize{Figure}}
\begin{center}
\includegraphics[scale=0.67]{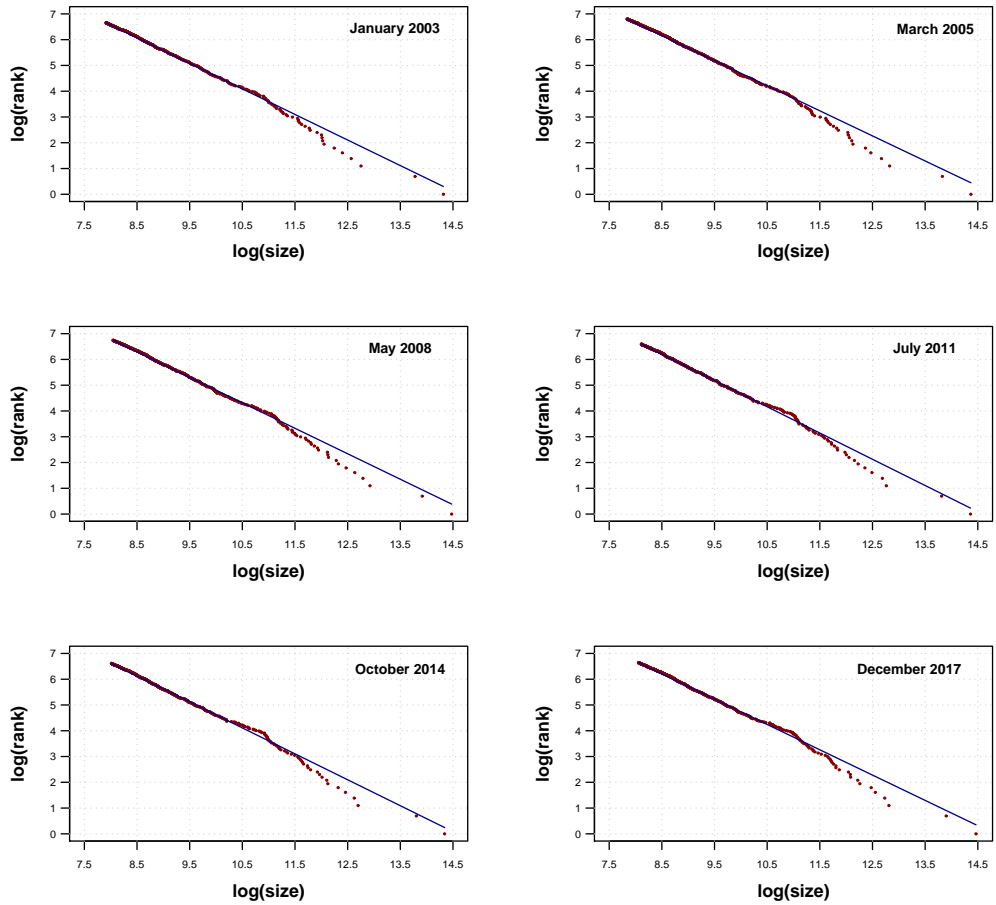}
\caption{\footnotesize{Rank-size plots, in the upper tail, of the complementary of the cumulative distribution
function, multiplied by $(n+1)$, of the power law model (solid lines) and the observed data, on log-log scale.
Data: Employment (workers affiliated) of the Spanish municipalities in January 2003, March 2005,
May 2008, July 2011, October 2014 and December 2017, published on the Spanish Social Security website.}}\label{fig04}
\end{center}
\end{figure}

Finally, as graphical model validation of the power law model in the upper tail of our data, we show in Figure \ref{fig04}
the rank-size plots (on log-log scale) of six different months, selected from the period Jan 2003 - Dec 2017.

\subsection{Modelling the whole range}
In this subsection, it is shown that the distribution of employment across municipalities in
Spain can be explained, in the whole range, with a Generalized Power Law (GPL) distribution.

For that reason, we choose a $GPL(II)$ model obtained from the real function
$g(z)=\alpha\left[\displaystyle\frac{\log (1+z)}{1+\log (1+z)}\right]^\beta$,
with $\mu=0$ ($Z=X/\sigma$), $\alpha,\sigma>0$ and $\beta>-1$,
defined in term of the survival function as follows (see Table \ref{tab4}),
\begin{equation}\label{gplcdf}
S_X(x)=\Pr(X>x)= \exp{\left[-\alpha\displaystyle\frac{\log^{\beta+1}(1+x/\sigma)}{[1+\log(1+x/\sigma)]^\beta}\right]}, \;x>0,
\end{equation}
and $S_X(x)=0$ if $x\leq 0$. Its probability density (pdf) function is given by,
\begin{equation}\label{gplpdf}\small
f_X(x)=
\displaystyle\frac{\alpha\left[\beta+1+\log\left(1+\frac{x}{\sigma}\right)\right]
\left[\log\left(1+\frac{x}{\sigma}\right)\right]^\beta}
{\sigma\left(1+\frac{x}{\sigma}\right)
\left[1+\log\left(1+\frac{x}{\sigma}\right)\right]^{\beta+1}}
\exp\left[-\alpha\displaystyle\frac{\left[\log\left(1+\frac{x}{\sigma}\right)\right]^{\beta+1}}
{\left[1+\log\left(1+\frac{x}{\sigma}\right)\right]^\beta}\right],\;x>0
\end{equation}
and $f(x)=0$ if $x\leq 0$.
This model includes Lomax (Pareto type II with $\mu=0$) distribution as particular case,
when $\beta=0$ (see Eq. \ref{eq15bis}). In addition, it belongs to the new subfamily defined by
the condition (\ref{eq6}):
$\displaystyle\lim_{z \to \infty}g(z)=\alpha>0$, thus it belongs to the maximum domain of attraction of the
Fr\'echet distribution, it is regularly varying at infinity with index $-\alpha<0$, it is right tail equivalent to a
Pareto (power law) model, and it is also a heavy-tailed distribution (see Remark \ref{rm2}).

We have fitted this $GPL(II)$ model in the whole range, in each of the
180 months considered (period Jan 2003 - Dec 2017), by using the maximum likelihood method of estimation.
It should be noted that the data are left-censored,
therefore the contribution to the maximum likelihood function
of the $r$ smallest municipalities, with less than or equal to 4 workers affiliated (reported as  "$<5$"),
would be $[1-S_X(4)]^r$, and the contribution of the rest of the municipalities, with 5 workers or more, would be
$\displaystyle\prod_{\substack{i=r+1}}^{N} f(x_i)$. The resulting log-likelihood function can be found in the Appendix.

Numerical computations for fitting the $GPL(II)$ distribution in the whole range, we have used the \verb|R| software
function \verb|optimx| \cite{Nash} with the limited memory quasi-Newton
L-BFGS-B algorithm \cite{Byrd}, by taking as initial values the corresponding
ones to the Lomax distribution with scale parameter equal to one:
$(\hat{\alpha}_0,\hat{\beta}_0,\hat{\sigma}_0)=(1,0,1)$.

\begin{figure}[htp]
\renewcommand{\figurename}{\footnotesize{Figure}}
\begin{center}
\includegraphics[scale=0.67]{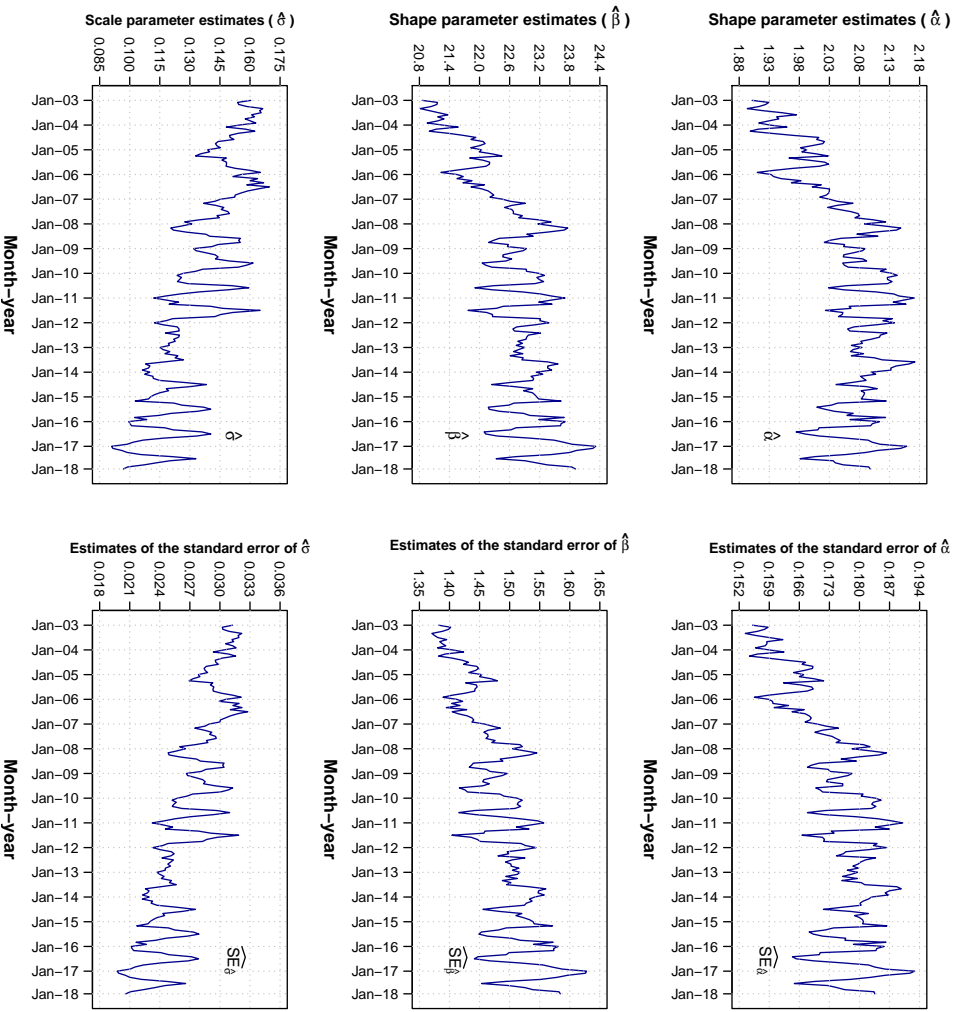}
\caption{\footnotesize{Parameter estimates for the chosen $GPL(II)$ model (Eq.\ref{gplcdf}), fitted
by maximum likelihood estimation in the whole range, and corresponding standard errors,
in each of the 180 monthly samples considered, from January 2003 to December 2017.}}\label{fig05}
\end{center}
\end{figure}

Figure \ref{fig05} shows the parameter estimates $(\hat{\alpha},\hat{\beta},\hat{\sigma})$ for the $GPL(II)$ model
fitted by maximum likelihood estimation, in each of the 180 month of the period considered (Jan 2003 - Dec 2017), and
the corresponding standard error $(\widehat{SE}_{\hat{\alpha}},\widehat{SE}_{\hat{\beta}},\widehat{SE}_{\hat{\sigma}})$.

\begin{figure}[p]
\renewcommand{\figurename}{\footnotesize{Figure}}
\begin{center}
\includegraphics[scale=0.44]{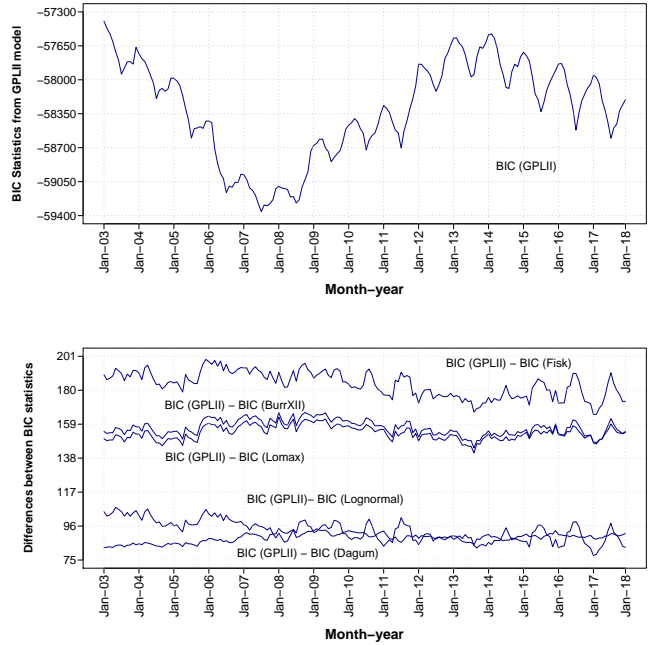}
\caption{\footnotesize{$BIC$ statistics for the chosen $GPL(II)$ model chosen (top part) and differences between the $BIC$ statistics for the $GPL(II)$ model and for Dagum, Lognormal, Lomax, Burr Type XII and Fisk distributions (bottom part), for each month of the investigation period. Positive differences mean that GPL model is preferable. }}\label{fig06}
\end{center}
\end{figure}

For comparison purposes, we have also fitted (by maximum likelihood, in the whole range,
with left-censored data, and in each of the 180 months considered) other well-known
models: (1) with three parameters, Dagum \cite{Dagum1975} and Burr type XII (Singh-Maddala)
distributions, and (2) with two parameters, Lognormal \cite{Aitchison1957}, Fisk (Log-logistic) and
Lomax (Pareto type II) distributions. We compared those distributions by using the Bayesian information
criterion ($BIC$), given by  the following expression \cite{Schwarz1978}
\begin{equation}\label{bic}
BIC=\log L-\frac{1}{2}d\log N
\end{equation}
where $\log L$ is the log-likelihood of the model evaluated at the maximum likelihood estimates,
$d$ is the number of parameters of that model, $N$ is the sample size. A model with largest value of the $BIC$ statistics is preferable. Top part of Figure \ref{fig06} shows  the value of $BIC$ statistics obtained from the chosen $GPL(II)$ distribution (Eq. \ref{gplcdf}), in each month of the period considered
(Jan 2003 - Dec 2017); bottom part illustrates the differences between the $BIC$ statistics for $GPL(II)$ model and
for the other five models (Dagum, Lognormal, Lomax, Burr Type XII and Fisk distributions). It is observable the differences of $BIC$ values are positive in each of the 180 months considered, which means that the $GPL(II)$ model is preferable to the other models. 

In addition, we have tested the goodness of fit of $GPL(II)$ model by bootstrap resampling for each of
the 180 months considered. First, by using the Kolmogorov-Smirnov ($KS$) statistic,
and then, by using the Anderson-Darling ($W_n^2$) statistic
. The latter test is more sensitive to deviation occurring in the tails \cite{Mason1983}. Taking into account that datasets are
left-censored, we have used the following computational formula for the ($W_n^2$) statistic:
$$W_N^2=-N-\frac{1}{N}\sum_{i=r+1}^{N}[2i-1][\log(F(x_{i}))+\log(1-F(x_{N+r+1-i}))]$$
where $F(x_{i};\hat{\alpha},\hat{\beta},\hat{\sigma})$ is the theoretical cumulative distribution function
(cdf) of the $GPL(II)$ model fitted by maximum likelihood in a sample value.

In both cases ($KS$ and $W_n^2$), and for each month, we have simulated 10,000 synthetic datasets
(left-censored for values less than or equal to $x_0=4$) by using its own
($\hat{\alpha},\hat{\beta},\hat{\sigma}$), and then we have calculated the corresponding
bootstrap $p$-values  as a fraction of the times that the empirical statistic is
better (less) than those 10,000 simulated statistics.
Those simulations were carried out on the Altamira Supercomputer at the Instituto de
F\'{\i}sica de Cantabria (IFCA).

\begin{figure}[p]
\renewcommand{\figurename}{\footnotesize{Figure}}
\begin{center}
\includegraphics[scale=0.44]{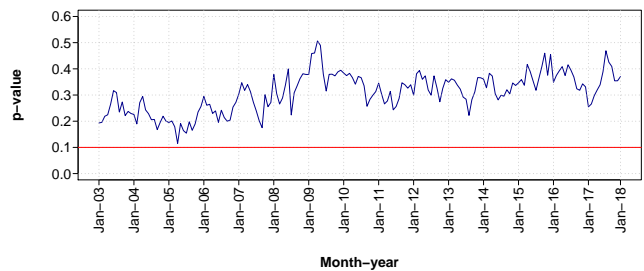}
\caption{
\footnotesize{Bootstrap $p$-values, obtained by using the Anderson-Darling ($W_n^2$) statistic,
in each of the 180 monthly samples considered, from January 2003 to December 2017
(as the $p$-values exceed the proposed significance level of 0.1,
there is not statistical evidence to reject {\it$H_0$: the data follow the chosen $GPL(II)$ model}, in the whole range
of the distribution).}}\label{fig07}
\end{center}
\end{figure}

We found that GPL(II) model is rejected at the proposed significance level by using the $KS$ statistics,
whereas by using the $W_n^2$ test statistics, GPL(II) model is not rejected at that
significance level.
Figure \ref{fig07} shows the bootstrap
$p$-values, obtained by using the $W_n^2$ statistics.
It is observable that the $p$-value exceeds 0.1 across all the months considered.

\begin{figure}[htp]
\renewcommand{\figurename}{\footnotesize{Figure}}
\begin{center}
\includegraphics[scale=0.67]{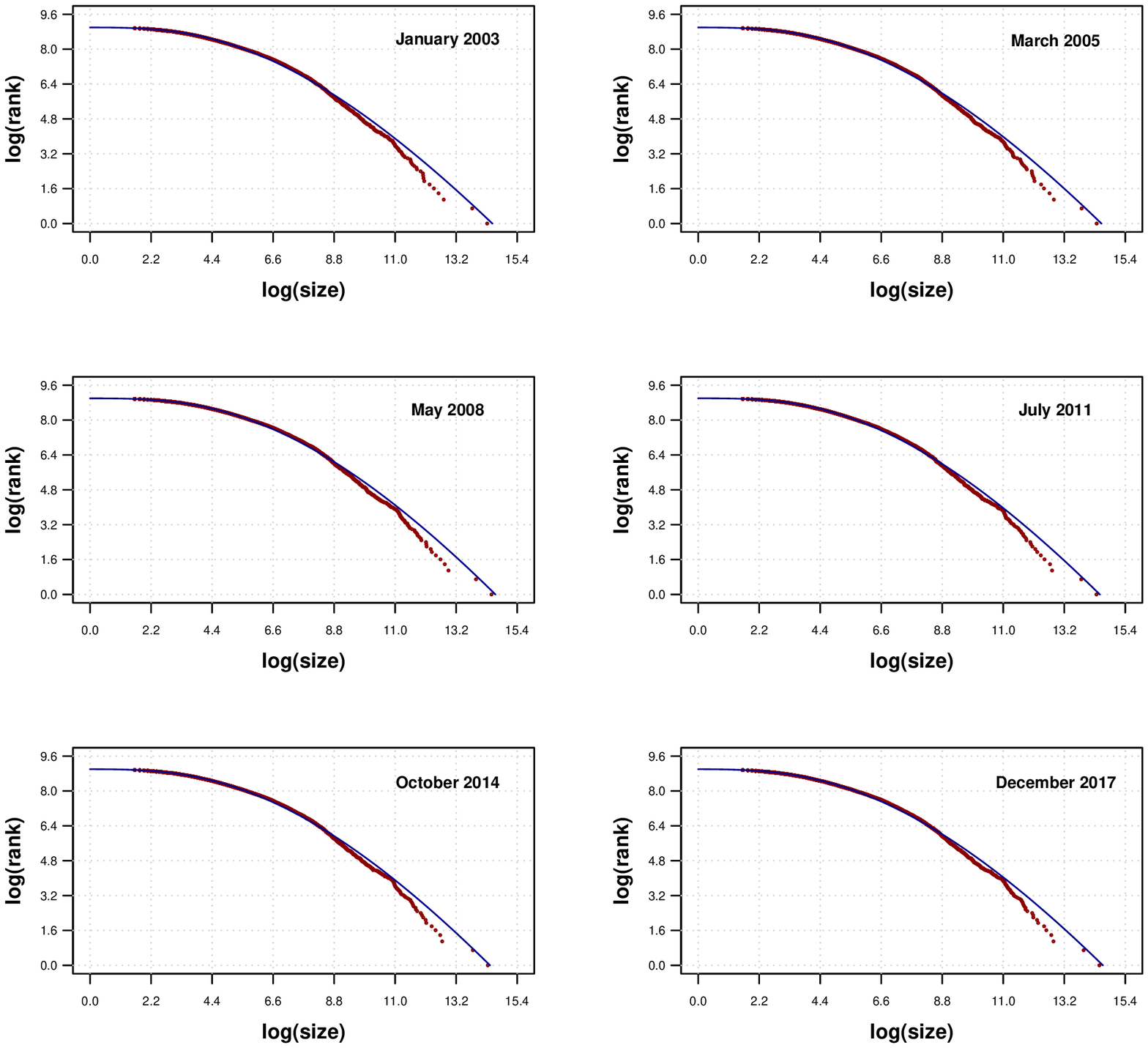}
\caption{\footnotesize{Rank-size plots, in the whole range, of the complementary of the cumulative
distribution function, multiplied by $(n+1)$,  of the $GPL(II)$ model (solid lines) and
the observed data (left-censored), on log-log scale.
Data: Employment (workers affiliated) of the municipalities in Spain in January 2003, March 2005,
May 2008, July 2011, October 2014 and December 2017, published on the Spanish Social Security
website.}}\label{fig08}
\end{center}
\end{figure}

Finally, as graphical model validation of this $GPL(II)$ model in the whole range of our data,
Figure \ref{fig08} displays the rank-size plots (on log-log scale) of six months,
selected from the period Jan 2003 - Dec 2017 considered.
It is appreciable an a slight deviation from the the theoretical model to the empirical data in the upper tail
of the distribution, corresponding to the biggest municipalities, with the exception of the two largest ones (Madrid and Barcelona) where the model fits reasonably well. Apart from this, the model provides an excellent fit in the rest of the distribution. 

\section{Conclusions} \label{se4}
In this paper, we analyzed the family of Generalized Power Law (GPL) distributions
(firstly introduced in \cite{Prieto2017}), and we derived a new subfamily of regularly varying distributions
at infinity with index $-\alpha<0$, that belongs to the Maximum
Domain of attraction of the Fréchet distribution. These heavy-tailed distributions are right tail equivalent to a Pareto (power law)
distribution.

The hierarchical structure of the GPL family that includes
the hierarchy of Pareto (power law) distributions as particular case was also deeply studied. 
In addition, we showed that well-known probabilistic models as PPS, Benini, Exponential, Rayleigh, Weibull, Gompertz, Singh-Maddala (Burr Type XII),
Fisk (Log-logistic) and Lomax are all members of the GPL family of distributions.

The nonlinear distribution of employment across municipalities in Spain was examined.
In this paper we have used datasets that include information about the number of workers (employed
or self-employed) affiliated to the Spanish Social Security system,
left-censored (municipalities with less than or equal to 4 workers are reported as  "$<5$"), on a monthly
basis, from January 2003 to December 2017. Our findings show that the distribution
of employment across municipalities followed a power law behavior in the upper tail of the distribution,
in each of the 180 months considered.

Finally, we showed that the distribution of that employment across municipalities can be modelled by means of a GPL model in the whole range. The latter provided a better fit to empirical data than other well-known
models in the literature such as Dagum, Lognormal, Lomax, Burr Type XII and Fisk distributions, in terms of Bayesian
information criterion ($BIC$). In addition, this GPL model was not be rejected at 0.1 significance level by using an Anderson-Darling
($W_n^2$) test statistic based on bootstrap resampling.

\section*{Acknowledgements}

Faustino Prieto acknowledges funding by the José Castillejo Programme (Grant number CAS17/00461,
Ministerio de Educación, Cultura y Deporte, Programa Estatal de Promoción de Talento y su Empleabilidad
en I+D+i, Subprograma Estatal de Movilidad, del Plan Estatal de Investigación Científica y Técnica y de
Innovación 2013-2016).
Faustino Prieto also acknowledges the Faculty of Business and Economics
and the Centre for Actuarial Studies at the University of Melbourne for their special support,
since part of this paper was written while Faustino Prieto was visiting the University of Melbourne
during the period July-September 2018.
Jos\'e Mar\'{\i}a Sarabia thanks to
Ministerio de Economía y Competitividad, project ECO2016-76203-C2-1-P, for partial support of this work.
The authors thankfully acknowledges the computer resources, technical expertise and
assistance provided by the Advanced Computing \& e-Science team at IFCA.
%We are grateful for the constructive suggestions provided by the reviewers, which have improved the paper.

\section*{Appendix}

\noindent {\bf Proof of Theorem \ref{th1}.} It should be noted that any function $g$,
with $\displaystyle\lim_{z \to \infty}g(z)=\alpha>0$,  is slowly varying at infinity:
$\displaystyle\lim_{z \to \infty}g(tz)/g(z)=1, \forall t>0$; then, we can check that
$\displaystyle\lim_{z \to \infty}\frac{S_Z(tz)}{S_Z(z)}=\displaystyle\lim_{z \to \infty}\frac{(1+tz)^{-g(tz)}}{(1+z)^{-g(z)}}=t^{-\alpha}, \forall t>0$.\\

\noindent {\bf Proof of Remark \ref{rm1}.} We can check that\\[0.5ex]
(a) $\displaystyle\lim_{z \to \infty}\frac{S_Z(z)}{G_Z(z)}=
\displaystyle\lim_{z \to \infty}\frac{(1+z)^{-g(z)}}{(1+z)^{-\alpha}}=1$, where  $G_Z(z)$ is (in this case)
the survival function of the Pareto type II distribution , $G_Z(z)=(1+z)^{-\alpha},\forall z>0$;\\[0.5ex]
(b) $\displaystyle\lim_{z \to \infty}\frac{S_Z(z)}{G_Z(z)}=
\displaystyle\lim_{z \to \infty}\frac{(1+z)^{-g(z)}}{e^{-\lambda z}}=\infty$,
where $G_Z(z)$ is (in this case) the survival function of the exponential distribution,
$G_Z(z)=e^{-\lambda z},\forall z>0$\\

\noindent {\bf Proof of Remark \ref{rm2}.} The proof of this result is straightforward by using a simple change
of variable in the proof of Theorem \ref{th1}. $Z=(X-\sigma)/\sigma$, $Z=(X-\mu)/\sigma$ and
$Z=[(X-\mu)/\sigma]^{1/\gamma}$ for $GPL(I)$, $GPL(II)$, $GPL(III)$ respectively.\\[-0.5ex]

\noindent {\bf Log-likelihood function}, $GPL(II)$ model (Eq.\ref{gplcdf}) with left-censored data:
\begin{equation*}\label{loglikelihood}
\begin{split}
\log\ell(\alpha,\beta,\sigma)&
=\log\left[[1-S_X(x_0)]^r\prod_{\substack{i=r+1}}^{N} f(x_i)\right]\\
&=(N-r)[\log(\alpha)-\log(\sigma)]-\sum_{\substack{i=r+1}}^{N}\log\left(1+x_i/\sigma\right)\\
&+\sum_{\substack{i=r+1}}^{N}\log\left[\beta+1+\log\left(1+x_i/\sigma\right)\right]
+\beta\sum_{\substack{i=r+1}}^{N}\log\left[\log\left(1+x_i/\sigma\right)\right]\\
&-(\beta+1)\sum_{\substack{i=r+1}}^{N}\log\left[1+\log\left(1+x_i/\sigma\right)\right]
-\alpha\sum_{\substack{i=r+1}}^{N}
\displaystyle\frac{[\log\left(1+x_i/\sigma\right)]^{\beta+1}}{[1+\log\left(1+x_i/\sigma\right)]^\beta}\\
&+r\log\left[1-\exp{\left[-\alpha\displaystyle\frac{[\log\left(1+x_0/\sigma\right)]^{\beta+1}}
{[1+\log(1+x_0/\sigma)]^\beta}\right]}     \right],
\end{split}
\end{equation*}
where $N$ is the sample size (included the left censored observations),
$r$ is the number of left censored observations,  $x_0=4$ is the censoring value,
and the maximum likelihood estimates of the unknown parameter vector
($\alpha$,$\beta$,$\sigma$) is the one that maximizes the log-likelihood function $\log\ell(\alpha,\beta,\sigma)$.

\bibliographystyle{plain}

\begin{thebibliography}{11}

\bibitem{Desmet2005}
Desmet K, Fafchamps M. Changes in the spatial concentration of employment across US counties:
a sectoral analysis 1972-2000. Journal of Economic Geography 2005; 5(3): 261-84.

\bibitem{Korpi2007}
Korpi M. Does size of local labour markets affect wage inequality? A rank-size rule of income distribution.
Journal of Economic Geography 2007; 8(2): 211-37.

\bibitem{Matano2011}
Matano A, Naticchioni P. Wage distribution and the spatial sorting of workers.
Journal of Economic Geography 2011; 12(2): 379-408.

\bibitem{Wojcik2015}
Wójcik D, MacDonald-Korth D. The British and the German financial sectors in the wake of the crisis: size,
structure and spatial concentration. Journal of Economic Geography 2015; 15(5): 1033-54.

\bibitem{McKenzie2016}
McKenzie FMH. Labour Force Mobility in the Australian Resources Industry. Springer; 2016.

\bibitem{Rampellini2016}
Rampellini K, Veenendaal B. Analysing the Spatial Distribution of Changing Labour Force Dynamics in the Pilbara.
In Labour Force Mobility in the Australian Resources Industry 2016; 29-58. Springer, Singapore.

\bibitem{Axtell2001}
Axtell RL. (2001). Zipf distribution of US firm sizes. Science 2001; 293(5536): 1818-20.

\bibitem{SS2018}
Spanish Social Security system. URL: http://www.seg-social.es/wps/\\
portal/wss/internet/EstadisticasPresupuestosEstudios/Estadisticas/\\
EST8/EST10/EST305/1836 ; [accessed 10 July 2018].

\bibitem{Prieto2017}
Prieto F, Sarabia JM. A generalization of the power law distribution with nonlinear exponent.
Communications in Nonlinear Science and Numerical Simulation 2017; 42: 215-28.

\bibitem{Ron1989}
Castillo E, Galambos J, Sarabia JM. The selection of the domain of attraction of an extreme value distribution from a set of data. In Extreme Value Theory 1989; 181-90. Springer New York.

\bibitem{Ron2005}
Castillo E, Hadi AS, Balakrishnan N, Sarabia JM. Extreme value and related models with applications in engineering and science. Hoboken, NJ: Wiley; 2005.

\bibitem{Seneta2006}
Seneta E. Regularly varying functions (Vol. 508). Springer; 2006.

\bibitem{Resnick2013}
Resnick SI. Extreme values, regular variation and point processes. Springer; 2013.

\bibitem{Kleiber2003}
Kleiber C, Kotz S. Statistical size distributions in economics and actuarial sciences (Vol. 470). John Wiley \& Sons; 2003.

\bibitem{Benini1905}
Benini R. I diagrammi a scala logaritmica (a proposito della graduazione per valore delle successioni ereditarie in Italia, Francia e Inghilterra). Giornale degli economisti 1905; 30: 222-31.

\bibitem{Sarabia2009}
Sarabia JM, Prieto F. The Pareto-positive stable distribution: A new descriptive model for city size data. Physica A: Statistical Mechanics and its Applications 2009; 388(19): 4179-91.

\bibitem{Guillen2011}
Guillen M, Prieto F, Sarabia JM. Modelling losses and locating the tail with the Pareto Positive Stable distribution. Insurance: Mathematics and Economics 2011; 49(3): 454-61.

\bibitem{Arnold2008}
Arnold BC. Pareto and generalized pareto distributions. In: Modeling income distributions and Lorenz curves 2008; 119-45. Springer New York.

\bibitem{Arnold2014}
Arnold BC. Univariate and multivariate Pareto models. Journal of Statistical Distributions and Applications 2014; 1(1): 1-16.

\bibitem{Arnold2015}
Arnold BC. Pareto distributions Second Edition. Chapman and Hall/CRC; 2015.

\bibitem{Lomax1954}
Lomax KS. Business failures; another example of the analysis of failure data. Journal of the American Statistical Association 1954; 49(268): 847-52.

\bibitem{Fisk1961}
Fisk PR. The graduation of income distributions. Econometrica: journal of the Econometric Society 1961: 171-85.

\bibitem{Burr1942}
Burr IW . Cumulative frequency functions. The Annals of Mathematical Statistics 1942; 13(2): 215-32.

\bibitem{Singh1976}
Singh S, Maddala G. A function for size distribution of incomes. Econometrica 1976; 44(5): 963-70.

\bibitem{Kotz2004}
Kotz S, Balakrishnan N, Johnson NL. Continuous multivariate distributions, models and applications. John Wiley \& Sons; 2004.

\bibitem{Bowley1920}
Bowley AL. Elements of statistics. New York: Charles Scribner's Sons; 1920.

\bibitem{Moors1988}
Moors JJA. A quantile alternative for kurtosis. The Statistician 1988; 37: 25-32.

\bibitem{Clauset2009}
Clauset A, Shalizi CR, Newman MEJ. Power-law distributions in empirical data.
SIAM Rev 2009; 51(4): 661-703.

\bibitem{Efron}
Efron B. Bootstrap methods: another look at the jackknife. Annals of Statistics 1979; 7(1): 1-26.

\bibitem{Babu}
Babu GJ, Rao CR. Goodness-of-fit tests when parameters are estimated. Sankhya 2004; 66: 63–74.

\bibitem{Hill1975}
Hill BM. A simple general approach to inference about the tail of a distribution. The Annals of Statistics 1975; 3(5): 1163-74.

\bibitem{Ioannides}
Ioannides YM, Overman HG.  Zipf's law for cities: an empirical examination. Regional Science and Urban Economics 2003; 33: 127-37.

\bibitem{Fujiwara}
Fujiwara Y, Guilmi CD, Aoyama H, Gallegati M, Souma W. Do Pareto-Zipf and Gibrat laws hold true? An analysis with European firms. Physica A: Statistical Mechanics and its Applications  2004; 335: 197-216.

\bibitem{Gabaix2016}
Gabaix X. Power laws in economics: an introduction. Journal of Economic Perspectives 2016; 30(1): 185-206.

\bibitem{Castillo2005}
Castillo E, Hadi AS, Balakrishnan N, Sarabia JM. Extreme value and related models with applications in engineering and science. John Wiley \& Sons; 2005.

\bibitem{Nash}
Nash JC, Varadhan R. Unifying optimization algorithms to aid software system users: optimx for R. Journal of Statistical Software 2011; 43(9): 1-14.

\bibitem{Byrd}
Byrd RH, Lu P, Nocedal J, Zhu CY. A limited memory algorithm for bound constrained optimization. SIAM Journal on Scientific Computing 1995; 16(5): 1190-208.

\bibitem{Dagum1975}
Dagum C. A model of income distribution and the conditions of existence of moments of finite order. In: Proceedings of the 40th session of the International Statistical Institute 1975; 46: 199-205.

\bibitem{Aitchison1957}
Aitchison J, Brown JAC. The lognormal distribution with special reference to its uses in economics, 1957.

\bibitem{Schwarz1978}
Schwarz G. Estimating the dimension of a model. Annals of Statistics 1978; 5: 461-64.

\bibitem{Mason1983}
Mason DM, Schuenemeyer JH. A modified Kolmogorov-Smirnov test
sensitive to tail alternatives. The Annals of Statistics 1983; 933-46.

\end{thebibliography}

\end{document}